\documentclass[useAMS,usenatbib]{mn2e}
\usepackage{graphicx}
\usepackage[usenames,dvipsnames]{color}
\usepackage{morefloats}
\usepackage{amsmath}

\title[X-ray polarimetry of Seyfert-1s]
      {Contribution of parsec-scale material onto the polarized X-ray spectrum of type-1 Seyfert galaxies}
\author[Marin et al.]
      {F. Marin$^1$\thanks{E-mail: frederic.marin@astro.unistra.fr},
       M. Dov{\v c}iak$^2$ and E. S. Kammoun$^3$\\
       $^1$Universit\'e de Strasbourg, CNRS, Observatoire Astronomique de Strasbourg, UMR 7550, F-67000 Strasbourg, France\\
       $^2$Astronomical Institute of the Academy of Sciences, Bo{\v c}n\'{\i} II 1401, CZ-14100 Prague, Czech Republic\\
       $^3$SISSA, via Bonomea 265, I-34135 Trieste, Italy}

\date{Accepted 2018 April 23;
      Received 2018 April 13;
      in original form 2018 March 9}

\pagerange{\pageref{firstpage}--\pageref{lastpage}}
\pubyear{2018}

\begin{document}

\maketitle

\begin{abstract}
Type-1 radio-quiet active galactic nuclei (AGN) are seen from the polar direction and offer a direct view of their
central X-ray engine. If most of X-ray photons have traveled from the primary source to the observer with
minimum light-matter interaction, a fraction of radiation is emitted at different directions and is reprocessed by 
the parsec-scale equatorial circumnuclear region or the polar outflows. It is still unclear how much the polarization 
expected from type-1 AGN is affected by radiation that have scattered on the distant AGN components. In this paper,
we examine the contribution of remote material onto the polarized X-ray spectrum of type-1 Seyfert galaxies using 
radiative transfer Monte Carlo codes. We find that the observed X-ray polarization strongly depends on the initial 
polarization emerging from the disk-corona system. For unpolarized and parallelly polarized photons (parallel to the disk), 
the contribution is negligible below 3~keV and tends to increase the polarization degree by up to one percentage points 
at higher energies, smoothing out the energy-dependent variations of the polarization angle. For perpendicularly polarized
corona photons, the addition of the circumnuclear scattered (parallel) component adds to the polarization above 10keV, 
decreases polarization below 10~keV and shifts the expected 90$^\circ$ rotation of the polarization angle to lower energies.
In conclusion, we found that simulations of Seyfert-1s that do not account for reprocessing on the parsec-scale equatorial 
and polar material are under- or over-estimating the X-ray polarization by 0.1 -- 1 percentage points. 
\end{abstract}


\label{firstpage}

\section{Introduction}
\label{Introduction}
In active galactic nuclei (AGN) that are not dominated by synchrotron processes in the form of failed or extended jets,
all observed X-ray fluxes originate from the vicinity of their central supermassive black hole (SMBH). It is
accepted that the SMBH accretion disk emits thermal ultraviolet photons that are reprocessed to the X-rays by Comptonization
processes in a hot corona lying above \citep{Haardt1991,Haardt1993} or maybe inside the disk \citep{Done2012}. The geometry,
spatial extension, composition and temperature of this corona have been investigated by many authors. The most recent 
observational constraints have been achieved in the hard X-ray spectrum of AGN by \textit{NuSTAR} which helped to improve our 
knowledge on the hot corona \citep{Fabian2015,Matt2015,Fabian2017,Tortosa2018}. Current constraints tend to picture the corona 
as compact, being only a few gravitational radii (r$_{\rm G}$) in size, with an anti-correlation between the coronal optical 
depth and the coronal temperature \citep{Tortosa2018}. Spectral, timing and reverberation studies also suggest a compact corona, 
most likely situated on the disk axis and located within 3 - 10 r$_{\rm G}$ from the central SMBH 
(\citealt{Zoghbi2010,Emmanoulopoulos2014,Gallo2015}, but see also \citet{Dovciak2016} for the minimum X-ray source size of 
the on-axis corona in AGN). X-ray microlensing analyses of lensed quasars also suggest that the X-ray emitting region is 
compact, having a half-light radius $\le$ 6 r$_{\rm G}$ \citep{Chartas2009,Mosquera2013,Reis2013}. However, the geometrical 
aspect of the corona is probably the most complicated aspect to study since spectroscopy has a limited sensitivity to the 
morphology of the emitting medium \citep{Wilkins2012}. Polarization, however, can distinguish more easily between
different geometries as the polarization of light is strongly influenced by morphological and composition variations
\citep{Dovciak2008,Schnittman2010,Marin2014}. The launch of the Imaging X-ray Polarimetry Explorer (\textit{IXPE}) in 2021 is
expected to revolutionize our understanding of the central engine of AGN by probing the geometry of 
the corona for the first time, among other fascinating topics \citep{Weisskopf2016}. 

\begin{figure}
    \includegraphics[trim = 0mm 70mm 0mm 10mm, clip, width=8.5cm]{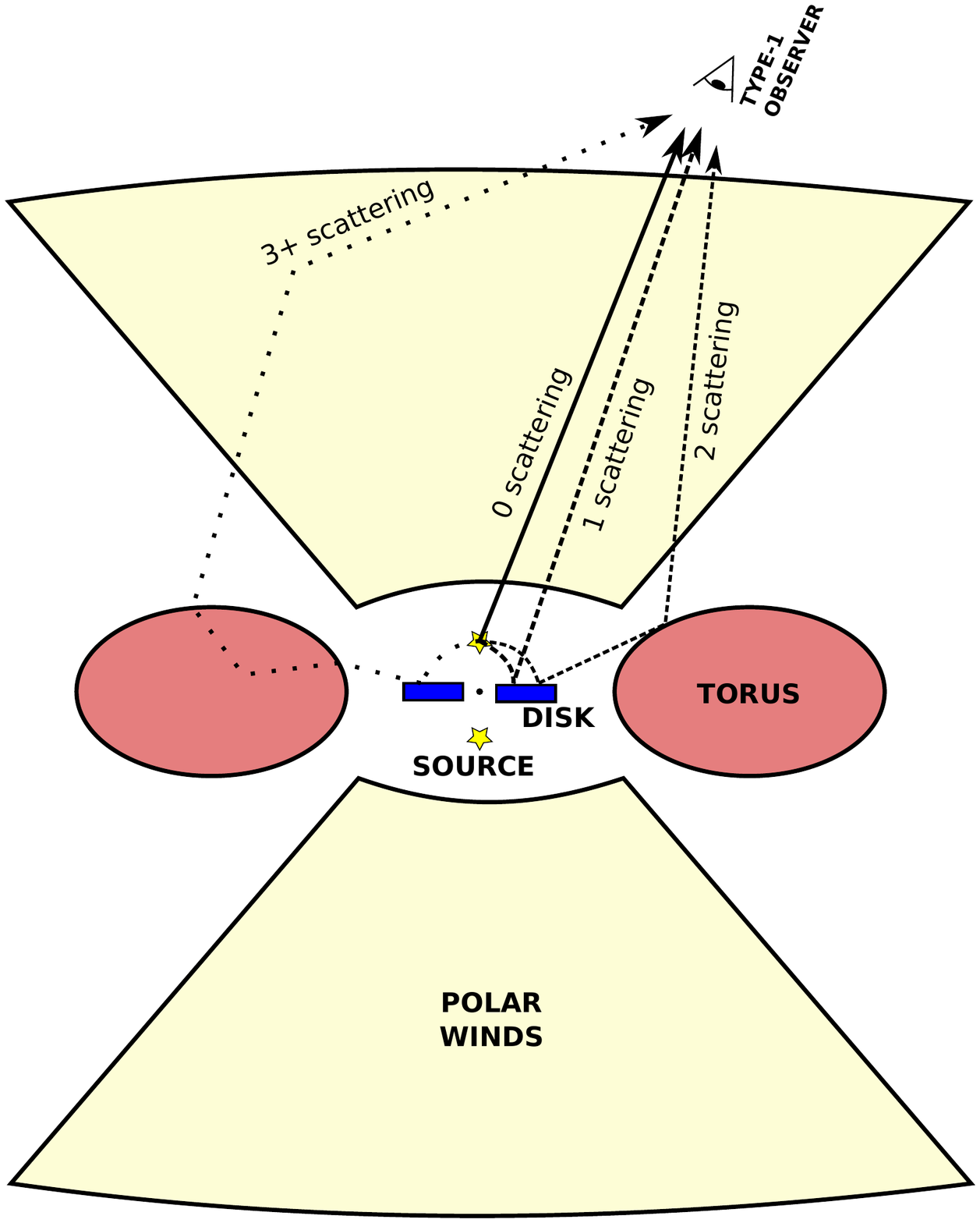}
    \caption{Artist representation of the AGN model. Scales
	    have been exaggerated for better visualization of 
	    the inner components. The point-like coronas 
	    are represented with yellow stars, the cold 
	    accretion disk is in blue, the gaseous torus 
	    in red and the polar outflows in primrose yellow. 
	    The photon trajectories are bend close to the 
	    central SMBH and radiation have multiple potential 
	    targets for interaction, depending upon the energy 
	    of the photon and Compton-thickness of the material.
	    The solid line represents the direct flux from the source 
	    (no scattering at all) and the other lines different 
	    possible radiation paths with different numbers of scattering.}
    \label{Fig:Scheme}
\end{figure}

In order to predict the degree and angle of X-ray polarization we expect from radio-quiet AGN, pro-active Monte Carlo 
simulations are necessary to explore in great details the radiative and reprocessing environment of SMBH 
\citep[see, e.g.,][]{Connors1980,Haardt1993a,Massaro1993,Dovciak2004a}. The central engine, consisting of a potential
well, an accretion disk and a lamp-post corona, is usually explored individually, disregarding any contribution from 
material above a few hundreds of gravitational radii. The processes occurring close to the singularity being complex 
due to special and general relativity, it is traditional to isolate the central AGN engine to estimate the X-ray 
polarization of type-1 AGN \citep[see, e.g.,][]{Dovciak2004a}. This, of course, cannot work for type-2 AGN, where the 
primary source is hidden behind a Compton-thick reservoir of cold matter situated at parsec-scales distances (see 
a modern AGN illustration in \citet{Marin2016}, or a simplified scheme in Fig.~\ref{Fig:Scheme}). The X-ray polarization 
emerging from type-2 AGN is strongly dominated by polar scattering inside the AGN outflows at soft X-ray energies,
but signatures of the disk-corona system may be detected in the hard band if the equatorial torus material has a 
Compton-thickness largely inferior to 10$^{25}$~at.cm$^{-2}$ \citep{Marin2018}. The parsec-scale AGN components are 
thus important to evaluate the X-ray properties of AGN, such as demonstrated by, e.g., \citet{Ghisellini1994} for 
spectroscopy and spectral energy distribution purposes. Yet, the contribution of those distant AGN components to
the X-ray polarization spectrum of type-1 Seyferts is not known.

This is the goal of our study: to couple the emission and scattering of X-ray photons by the disk-corona system 
governed by strong gravity effects to absorption, re-emission and scattering processes happening at parsec-scale 
distances from the potential well. This investigation is a follow up of the work from \citet{Marin2018}, where the 
same model and same codes were applied to estimate the X-ray polarization signal of type-2 AGN. In this paper, we 
check whether the parsec-scale torus and polar winds have an impact on the expected polarization of type-1 
AGN. If so, we evaluate how much of the signal is diluted or enhanced. To do so, we briefly remind the reader 
about the model and Monte Carlo codes used in our simulations in Sect.~\ref{Modeling}. We then present our 
results in Sect.~\ref{Results}, where different initial polarization state of the corona are investigated. In 
Sect.~\ref{Discussion} we explore how much the Compton-thickness of the outflows, and circumnuclear region
influence our results. We conclude our paper in Sect.~\ref{Conclusions} by quantifying the excess or deficiency 
of X-ray polarization estimated by previous type-1 AGN simulations.

%---------------------------------------------------------------------------------------------------%
\section{Modeling}
\label{Modeling}
Our modeling is exactly the same as previously described in details in \citet{Marin2018}. The sole difference is 
that we are looking at a typical type-1 inclination ($\sim$ 20$^\circ$ from the polar axis of the model) rather 
than at a type-2 orientation. In the following, we remind the reader about the Monte Carlo codes used and the 
geometry and composition of the AGN model.

\subsection{The radiative transfer codes}
\label{Modeling:code}
To compute the radiative transfer of photons emitted from a spherical corona situated above the accretion disk 
in a lamp-post geometry, we used the KY code presented in \citet{Dovciak2004}. KY is a relativistic ray-tracing 
code that computes the time evolution of a local spectrum seen by a distant observer. The code includes the 
full set of Stokes parameters to compute polarization, accounts for a variety of inclinations, black hole 
masses, corona locations, and potential orbiting clouds. Reprocessing in the disk is calculated with the 
Monte Carlo multi-scattering code NOAR \citep{Dumont2000}, that computes the reflected flux including the iron fluorescent 
K$\alpha$ and K$\beta$ lines. The single-scattering approximation \citep{Chandrasekhar1960} is used for the local 
polarization of the reflected continuum component while the line flux is assumed to be unpolarized.

Once the radiation field emerging from the disk-corona system computed, we inject the photons into the 
Monte Carlo code STOKES. Designed by \citet{Goosmann2007} and developed by \citet{Marin2012,Marin2015} and 
\citet{Rojas2017}, STOKES is a Monte Carlo code that simulates light reprocessing in any three-dimensional environment.
The code can compute the effect of multiple scattering and radiative coupling between a large number of media, 
from the near-infrared to the hard X-ray band. All the necessary absorption, re-emission and scattering physics 
is included in the code to cover such a large waveband \citep{Marin2014a}. Escaping photons are finally recorded
by a spherical web of virtual detectors at all polar and azimuthal directions. Using the output of KY as an input 
for STOKES, we are thus able to provide a consistent picture of the X-ray photons that emerge from the few 
gravitational radii around the central SMBH to reach the parsec-scale regions that constitute the external envelope
of AGN.

\subsection{The AGN model}
\label{Modeling:AGN}

\begin{figure}
    \includegraphics[trim = 0mm 0mm 0mm 0mm, clip, width=8.5cm]{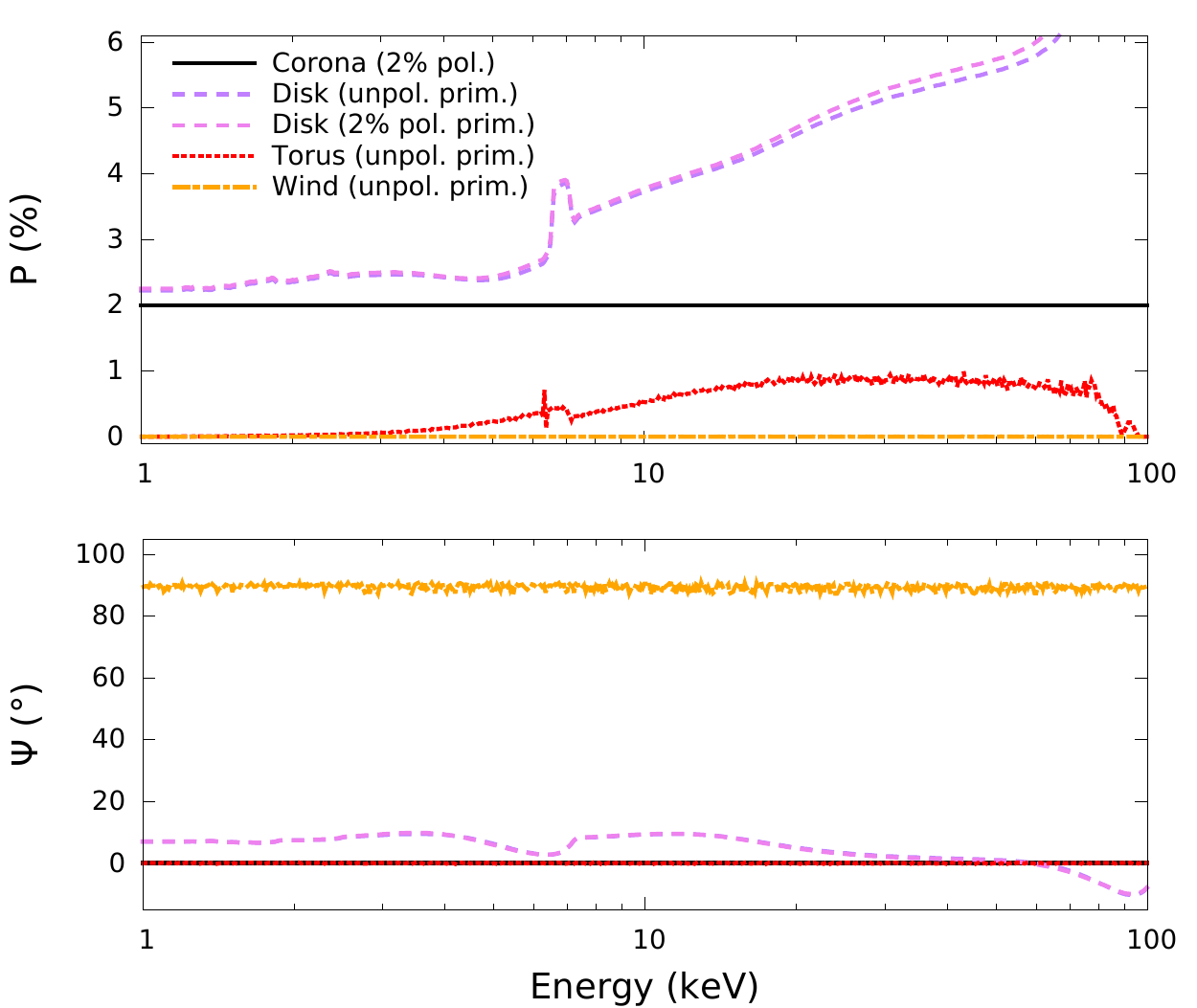}
    \caption{X-ray polarization degree (top panel) and position angle (bottom panel)
	     from the different components of our AGN model. The inclination of the observer 
	     is 20$^\circ$ and the photon source is either unpolarized or polarized at 2\% 
	     level. Solid line (black): polarization from the photons originating from the 
	     corona (no scatterings); dashed line (purple): polarization scattered 
	     from the disk using an unpolarized primary; dashed line (violet): polarization 
	     scattered from the disk using a 2\% polarized primary; dotted line (red): 
	     polarization scattered from the torus using an unpolarized primary; and 
	     dot-dashed line (orange): polarization scattered from the polar winds 
	     using an unpolarized primary.}
    \label{Fig:Decomposition}
\end{figure}

\begin{figure}
    \includegraphics[trim = 0mm 0mm 0mm 0mm, clip, width=8.5cm]{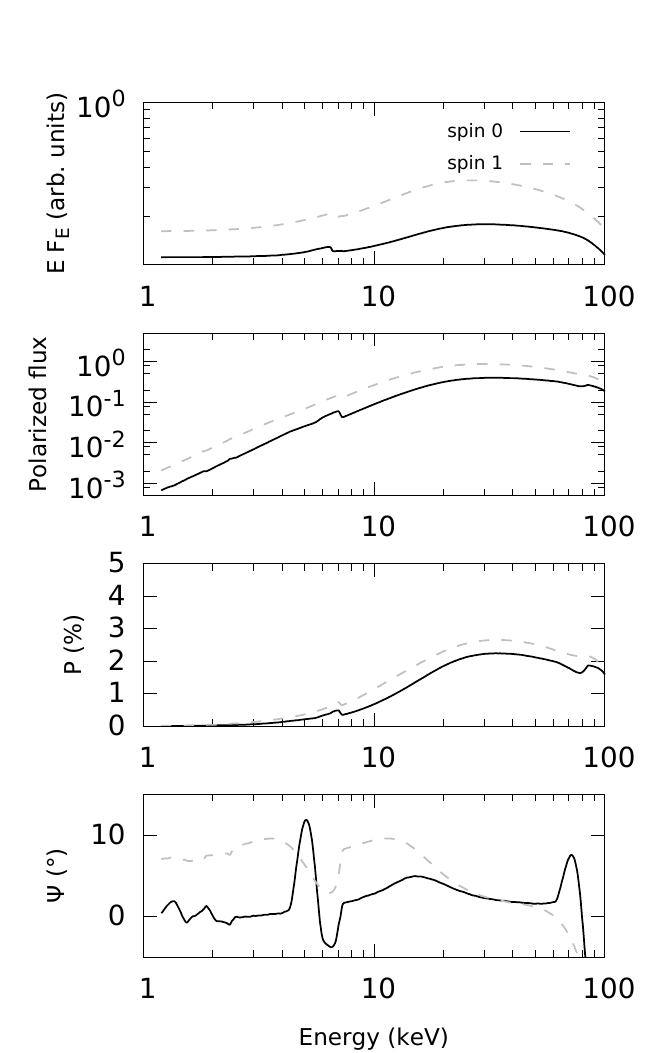}
    \caption{X-ray flux (F$_{\rm E}$ is energy flux at energy E), polarized flux, polarization degree 
	    $P$ and polarization position angle $\Psi$ seen by an observer at infinity, resulting from an 
	    elevated point-like corona that irradiates an accretion disk inclined by 20$^\circ$. 
	    The source is unpolarized. The variation in $P$ and $\Psi$ are due to general 
	    relativistic effects that will induce a parallel transport of the polarization 
	    angle along geodesics, plus the scattering and re-emission of photons from the 
	    cold accretion matter. Two flavors of black holes are shown: a non-spinning 
	    Schwarzschild black hole (black solid line) and a maximally spinning Kerr 
	    black hole (gray dashed line).}
    \label{Fig:Initial_NOTpolarization}
\end{figure}

Our AGN model is presented in Fig.~\ref{Fig:Scheme}. It consists of a central SMBH of $\sim$ 10$^8$ solar masses 
that represents the median value of the black hole mass distribution observed in large radio-quiet Seyfert catalogs
by \citet{Woo2002}. The dimensionless spin parameter of the black hole is either fixed to 0 (non-rotating 
Schwarzschild case) or 1 (maximally rotating Kerr case). Around the potential well we set a geometrically 
thin, optically thick accretion disk. Its inner radius is fixed by the spin value and its outer radius extends
up to 1000 gravitational radii, where strong gravity effects are no longer necessary. Since we do not consider 
super-Eddington flows that would required slim/flared disks \citep{Sadowski2011}, our disk is modeled using the 
usual slab geometry \citep[e.g.,][]{Dovciak2004}. The standard thin disk is uniformly filled with cold material, 
a valid scenario as long as the incident X-ray flux is smaller than or comparable with the soft thermal flux 
generated intrinsically in the disk \citep{Nayakshin2001}. New tables for moderately and highly ionized disks 
are currently computed (Goosmann et al., in prep). The hot corona in our model is set along the rotation axis of
the disk at a height of 3 gravitational radii. The point-like corona emits a power-law X-ray spectrum with a 
photon index $\Gamma$ = 2 and initial photons can have any intrinsic polarization.

Surrounding the central engine, an obscuring circumnuclear region usually refereed to as the ``torus'' prevents
radiation from escaping the equatorial plane. The half-opening angle of the torus is set to 60$^\circ$ 
from the polar axis, so the line-of-sight of a polar observer is unobscured by the gaseous medium. The inner 
radius of the cold material is fixed to 0.01~pc from the center of the model according to reverberation mapping 
data \citep{Suganuma2006,Vazquez2014}, and extends up to 5~pc. We also fixed the hydrogen column density along 
equatorial plane to 10$^{24}$~at.cm$^{-2}$. The torus is thus at the limit between Compton-thin and Compton-thick 
categories and can be representative of both AGN classes \citep{Georgakakis2017}. Finally, along the pole, we 
added outflowing winds collimated by the torus funnel. The winds extend 60$^\circ$ from the polar axis and
are composed of neutral gas in a Compton-thin regime (10$^{21}$~at.cm$^{-2}$). Those winds correspond to the 
Narrow Line Region (NLR) that extends over several kilo-parsec in nearby AGN before mixing with the interstellar 
medium \citep{Wang2016}. The NLR is characterized with little absorption features in the X-ray band and can be 
simply modeled using cold gaseous material (\textit{ibid.}). More ionized material, such as the warm absorber region 
\citep{Reynolds1995,Porquet2000} or the ultra-fast outflows \citep{Tombesi2010,Tombesi2011,Tombesi2012} are not 
explored here since they would not strongly contribute to the expected X-ray polarization continuum of AGN 
\citep{Marin2018}. We thus simplify our model by accounting for only one large polar region. The wind base is 
located at a radial distance of 0.1~pc from the center of the model \citep{Taniguchi1999} and extends up to 25~pc, 
before mixing with the interstellar medium. 

In total, there are three different targets for radiation to interact with: the accretion disk, the torus and the 
outflows. The resulting polarization emerging from each isolated component is presented in Fig.~\ref{Fig:Decomposition}.
If the primary source of photons (the corona) has no intrinsic polarization, it will only dilute the observed polarized 
flux. However, if it is intrinsically polarized by inverse Compton scattering of disk photons, the corona will also play 
a role in determining the polarization at infinity. In our model, we see that the direct flux of photons coming from 
the corona is polarized at a constant level that is independent of energy. Not all primary photons directly reach the 
observer; a substantial fraction of corona radiation is bent towards the disk by relativistic effects. Photons scattered 
by the disk keep memory of their initial polarization but, according to the scattering position with respect to the 
potential well and the energy of the photon, their polarization degree and angle will change. We show two different 
cases where the initial polarization is either 0 or 2\%. Interestingly, the two scattering-induced polarization levels 
are very similar. Our choice to use a 2\% polarized primary is driven by the fact that a similar level of polarization 
(for a viewing angle of 20$^\circ$) is produced by reflection from the disk. This allows us to model the polarization 
of a central engine that is not entirely dominated by the polarization of the corona. We also point out that the Compton
hump polarization, resulting from corona-disk scattering, can be quite high in both cases (see, e.g., Fig.~6 in 
\citealt{Dovciak2011}), and it strongly depends on the height of the primary source, the inclination of the observer,
and the spin of the black hole. Looking at the polarization produced by the parsec-scale components, we see that 
scattering of unpolarized photons by the torus produces a polarization degree of 0.1 -- 1.2\% with a constant polarization
angle of 0$^\circ$. On the contrary, forward scattering in the polar winds produces almost no polarization and its 
position angle is perpendicular to the polarization angle from the torus. For both spectra, we used an unpolarized 
primary to better isolate the true polarization resulting from scattering onto the parsec-scale components. 

Polarization being a vectorial quantity, it is not possible to simply add the contribution of each element 
to create the final polarized spectrum. The polarization emerging from a model where several components are coupled 
is thus not easily estimated by linear scaling or approximations without prior knwoledge of the total AGN polarization. 
The impact of radiative coupling between the various AGN constituents is precisely what we intend to explore in this 
paper. For all the simulations we present, the observer's inclination is fixed to a typical type-1 value of 20$^\circ$.

\subsection{Polarization of the primary source}
\label{Modeling:Polarization}

Our primary source is a point-like corona above the accretion disk. Its physical size and composition is not 
relevant to this paper since, in the following, all our AGN spectra will be compared to the spectra emerging from the 
disk-corona system. The differences we will detect can be extrapolated for all different corona properties. The corona 
itself may, however, emit different flavors of photons. Either the interaction of ultraviolet photons with the hot 
electrons produces unpolarized X-ray photons, or the outgoing photons are carrying a substantial polarization degree 
with a given polarization position angle. Detecting those polarization levels is extremely important to determine 
the composition, temperature and geometry of the corona and it will be an important target for \textit{IXPE}
\citep{Marin2014,Weisskopf2016}.

\begin{figure*}
    \includegraphics[trim = 0mm 0mm 0mm 0mm, clip, width=8.5cm]{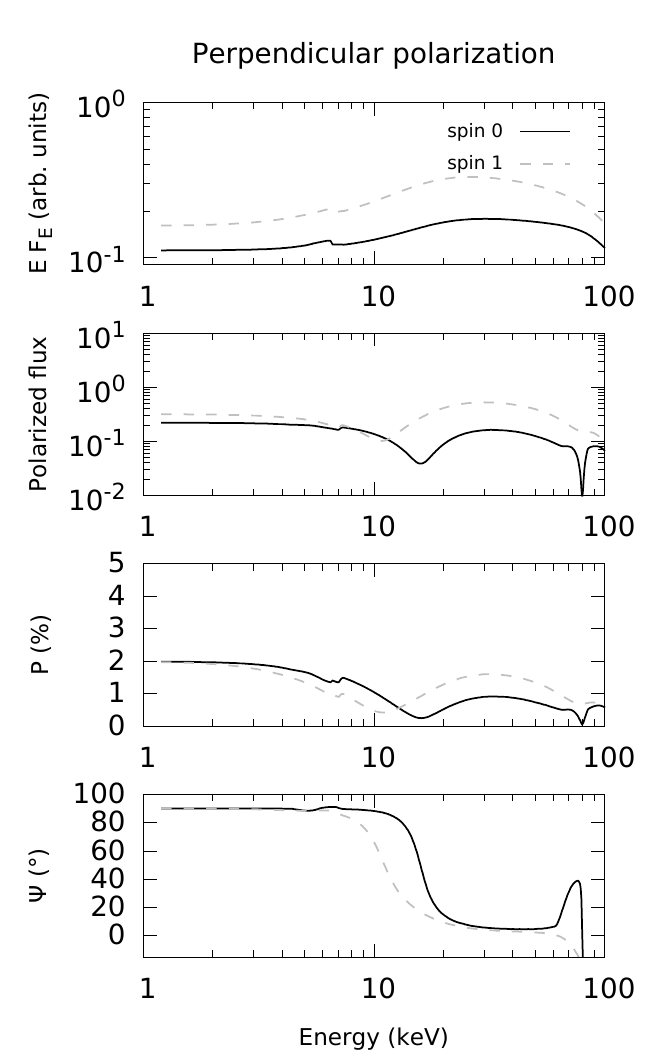}  
    \includegraphics[trim = 0mm 0mm 0mm 0mm, clip, width=8.5cm]{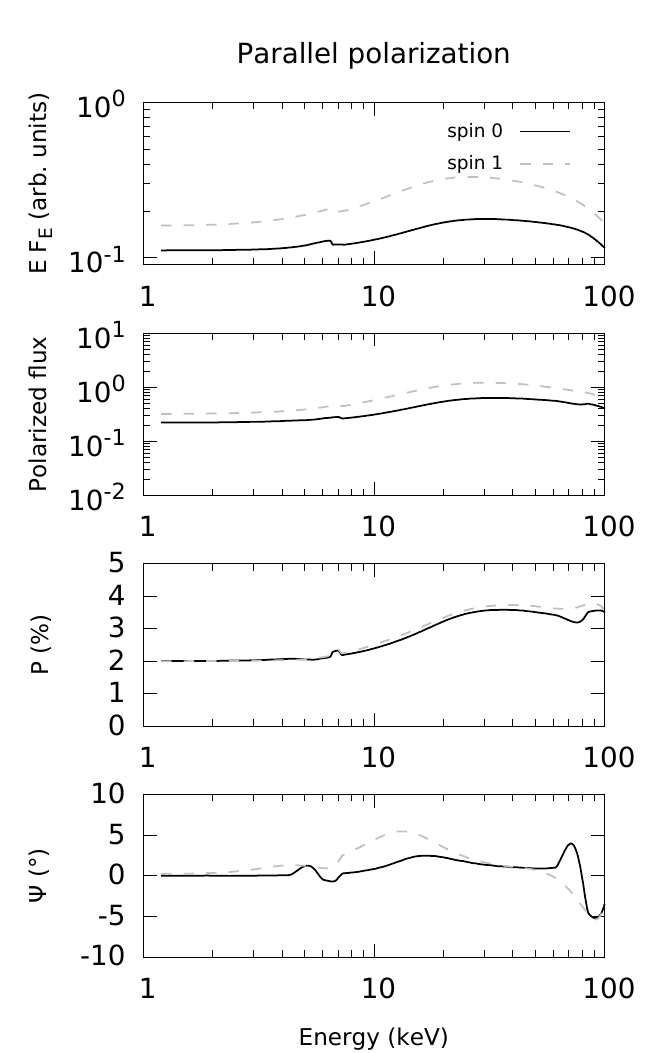}
    \caption{X-ray flux (F$_{\rm E}$ is energy flux at energy E), polarized flux, polarization degree 
	    $P$ and polarization position angle $\Psi$ seen by an observer at infinity, resulting from an 
	    elevated point-like corona that irradiates an accretion disk inclined by 20$^\circ$.
	    The initial polarization is set to 2\% with a parallel or a perpendicular polarization 
	    angle (right and left column, respectively). The variation in $P$ and $\Psi$ 
	    are due to general relativistic effects that will induce a parallel transport 
	    of the polarization angle along geodesics, plus the scattering and re-emission 
	    of photons from the cold accretion matter. Two flavors of black holes are 
	    shown: a non-spinning Schwarzschild black hole (black solid line) and a maximally 
	    spinning Kerr black hole (gray dashed line).}
    \label{Fig:Initial_polarization_20deg}
\end{figure*}

In this paper, similarly to \citet{Marin2018}, we investigate three different coronal polarization. In 
Fig.~\ref{Fig:Initial_NOTpolarization}, the input photons are not polarized. Coronal radiation travels along photons
null geodesics before escaping towards the observer or impacting the disk. The local polarization is computed and 
the photon can either be absorbed, scatter several times in the disk, or escape. The strong gravity effects 
influence the polarization position angle of radiation as it travels close to the potential well \citep{Connors1980}. 
The differences we see in Fig.~\ref{Fig:Initial_NOTpolarization} between a Schwarzschild and a Kerr model is due 
to the radius of the innermost stable circular orbit that is six times smaller in the latter case. The receding and 
approaching parts of the disk contributing differently to the sum of polarization also explains the differences 
in terms of polarization degree $P$ and polarization angle $\Psi$. Details of the energy-dependent variations 
of the spectra are extensively given in \citet{Marin2018}.

In Fig.~\ref{Fig:Initial_polarization_20deg}, we show the same modeling of the disk-corona system but with 
a 2\% polarized primary. While the true initial polarization of X-ray coronas is still unknown, 
as it depends on its geometry, optical depth, temperature and the inclination of the whole system, recent 
numerical simulations favor low polarizations degrees, of the order of a few percents \citep{Schnittman2010,Matt2018}.
As explained in the previous section, we choose a 2\% primary polarization, a value that has the particularity 
of being close to that of the polarization produced by the disk at an inclination of 20$^\circ$. Thus, we can
investigate the specific case where the polarization signal from the corona is not dominating the polarization 
from the disk. The left column shows an initial polarization position angle of the primary source 
perpendicular to the disk surface, while the right column is for parallel polarization (polarization angle 
parallel to the disk). We define a parallel polarization angle to be 0$^\circ$ and a perpendicular polarization
angle to be 90$^\circ$. We caution the reader that this is a different convention than what was used in 
\citet{Marin2018}. We changed the convention in order to be in agreement with other polarization codes. The
most striking difference between a polarized and an unpolarized source is the degree of polarization. At low 
energies, where photo-absorption in the disk dominates, the degree of polarization is the same as the corona 
input polarization. However, once the disk starts to reflect X-ray light, the polarization either increases 
or decreases depending on the polarization angle of input radiation. Scattering inside the disk tends to give
parallel polarization angles, so if the input photons are already parallelly polarized the resulting polarization
degree increases. Otherwise there is orthogonality between the two vectorial components, decreasing the net 
polarization degree. This is the reason why the polarization angle rotates between 90$^\circ$ and 0$^\circ$ 
between 10~keV and 20~keV. Finally, the strong feature appearing at very high energies is a numerical artifact 
caused by the truncated power-law and the energy shift of photons that affect energies higher than 70~keV 
and create empty bins.

%---------------------------------------------------------------------------------------------------%
\section{Results}
\label{Results}
The results presented in Figs.~\ref{Fig:Initial_NOTpolarization} and \ref{Fig:Initial_polarization_20deg} only 
account for the central engine. The contribution of the parsec-scale AGN components is disregarded. Those results 
are typically used to predict the degree and angle of polarization a future X-ray polarimetric satellite will 
observe in the case of type-1 AGN after the addition of Galactic absorption along the line-of-sight. In this section, 
we include the central AGN engine inside a larger model with a torus and polar winds (see Fig.~\ref{Fig:Scheme}). 
By doing so, we can estimate how different the polarimetric results are once the contribution of all AGN parts
are accounted for.

\subsection{Type-1 spectra with an unpolarized primary source}
\label{Results:UNPOL}

\begin{figure*}
    \includegraphics[trim = 0mm 0mm 0mm 0mm, clip, width=18cm]{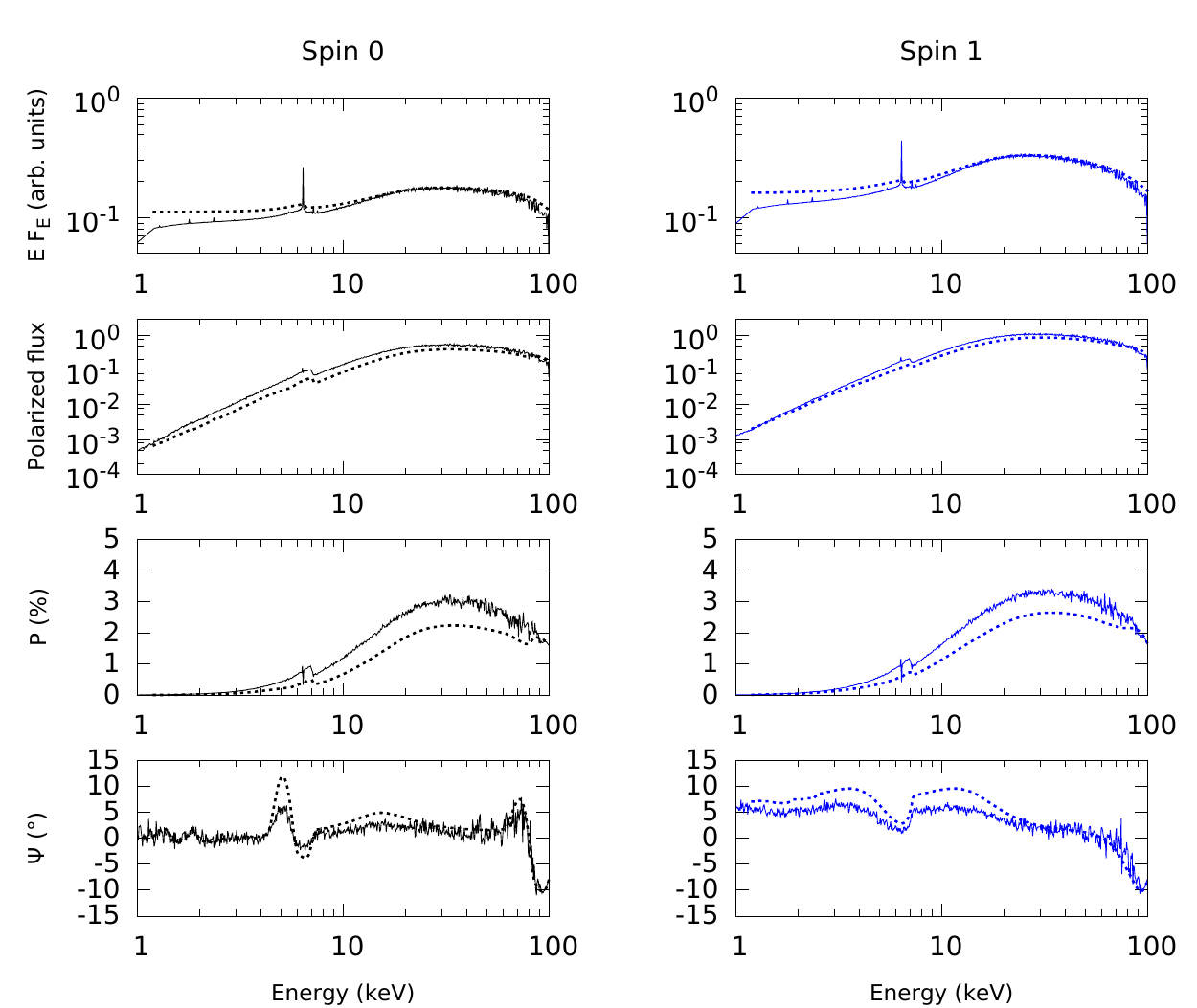}
    \caption{X-ray flux (F$_{\rm E}$ is energy flux at energy E), polarized flux, 
	    polarization degree and polarization position angle for a type-1 AGN
	    with a Compton-thick torus (n$_{\rm H_{torus}}$ = 10$^{24}$~at.cm$^{-2}$) 
	    and Compton-thin absorbing polar winds (n$_{\rm H_{wind}}$ = 10$^{21}$~at.cm$^{-2}$).
	    See text for additional details about the model components. The input 
	    corona spectrum is unpolarized and shown in dotted line.}
    \label{Fig:Unpolarized_primary}
\end{figure*}

We start our investigation with a model where the primary source emits photons that are randomly polarized 
(the resulting net polarization is thus null). We see from Fig.~\ref{Fig:Unpolarized_primary} (Schwarzschild 
case, left column) that the X-ray intensity spectrum of the AGN model is not strongly affected by the presence
of the torus and winds. There is slightly more absorption in the soft X-ray band with respect to the input 
disk-corona spectrum (in dashed line). This extra absorption is similar to what we observe when we take into 
account Galactic absorption \citep{Zamorani1988} and is due to the Compton-thin polar winds. The very soft 
X-ray photons are absorbed by the cold material while higher energy photons pass through without interacting. 
We also detect the presence of fluorescence lines that originate from the winds and torus, with a narrow 
Fe K$\alpha$ dominating the spectrum at 6.4~keV. This narrow component is an almost ubiquitous feature in 
the X-ray spectra of AGN and, in our case, mainly originates from the neutral material located in the molecular 
torus \citep[see also][]{Ricci2014}. There are no absorption lines since we do not model the warm absorber region.
The X-ray polarized spectrum (second panel from the top) appears to be very similar to the input polarized spectrum, 
being marginally higher due to extra scattering on the parsec-scale components. The polarized flux being the 
multiplication of the intensity spectrum and the polarization degree, it is then logical to observe that the input 
corona-disk system polarization is lower than the final full AGN model polarization degree $P$. If the difference 
is negligible in the soft band ($E <$ 3~keV), the AGN polarization degree is twice higher in the 3 -- 10~keV band,
increasing by 0.1 -- 0.5 percentage points. The difference is even higher in the Compton hump where the polarization
degree increases from 2\% to 3\%. Multiple scatterings between the disk-corona system and the parsec-scale components
play a not-so-negligible role in determining the true X-ray polarization expected from type-1 AGN. The main contributor
to this additional polarization is the torus, which produces an additional parallelly polarized component that affects 
the whole energy band. The differences in terms of polarization position angle $\Psi$ are less important as the 
energy-dependent variations of $\Psi$ remain the same. Only the intensity of the variations are smoothed out since 
scattering by the equatorial parsec-scale component tend to fix the polarization angle to 0$^\circ$.

The case is very similar for a Kerr black hole (Fig.~\ref{Fig:Unpolarized_primary}, right column), 
with the exception of the intensity spectrum that shows higher fluxes but with the same absorption levels 
in the soft X-ray band. The principal method to distinguish between a non-rotating and a maximally rotating 
black hole relies on the energy-dependent variations of $\Psi$, which are conserved despite the addition 
of the torus and polar outflows. It is thus safe to say that adding parsec-scale components do not modify 
past predictions if the coronal photons are unpolarized. The parsec-scale components only uplift previous 
estimations of the X-ray polarization degree we expect, which is more favorable for observations.

\subsection{Type-1 spectra with a (2\% parallel) polarized primary source}
\label{Results:POL_PARA}

\begin{figure*}
    \includegraphics[trim = 0mm 0mm 0mm 0mm, clip, width=18cm]{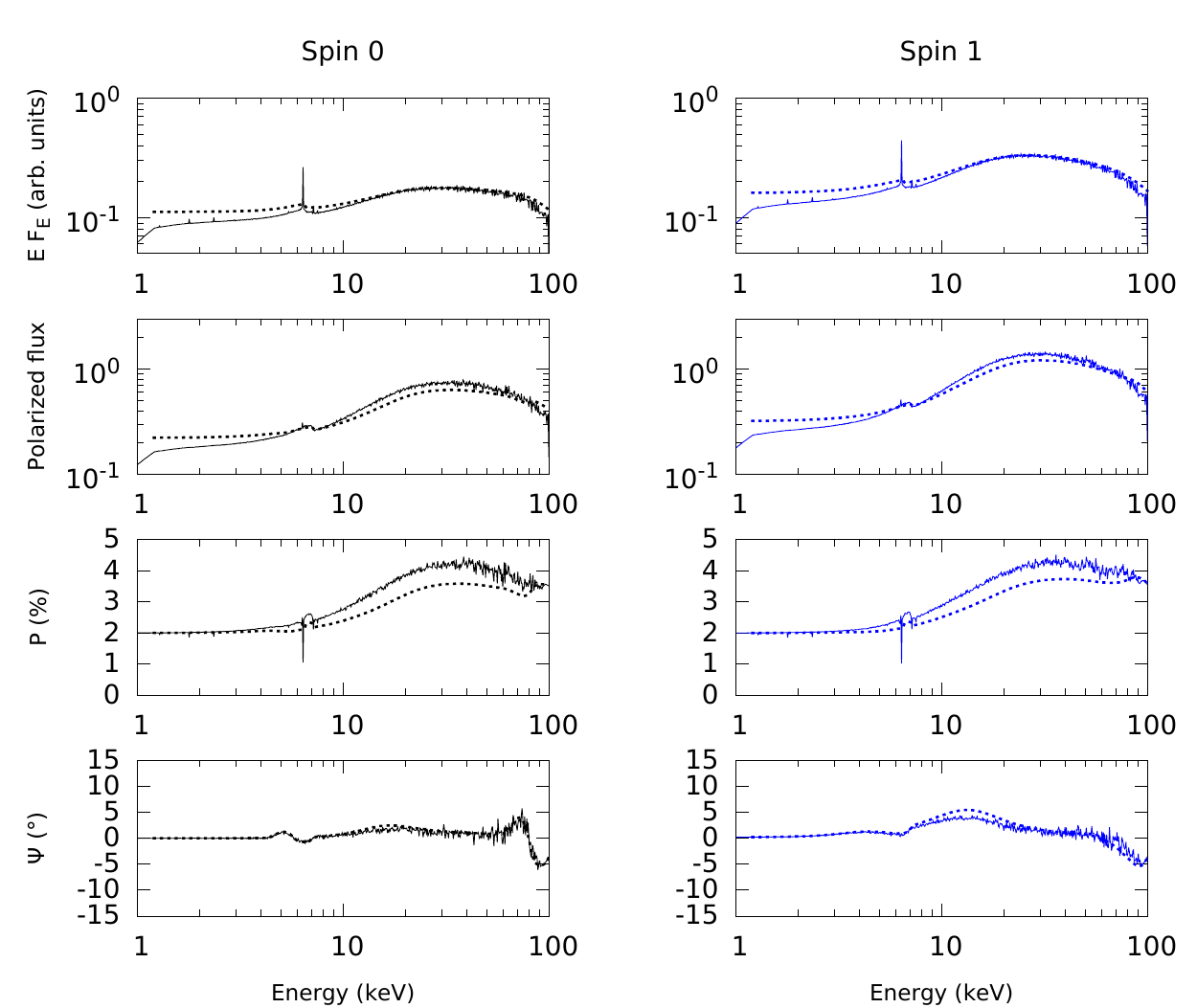}
    \caption{X-ray flux (F$_{\rm E}$ is energy flux at energy E), polarized flux, 
	    polarization degree and polarization position angle for a type-1 AGN
	    with a Compton-thick torus (n$_{\rm H_{torus}}$ = 10$^{24}$~at.cm$^{-2}$) 
	    and Compton-thin absorbing polar winds (n$_{\rm H_{wind}}$ = 10$^{21}$~at.cm$^{-2}$).
	    The input corona spectrum is 2\% parallelly polarized and shown in dotted line.}
    \label{Fig:Polarized_para}
\end{figure*}

We continue our tests by setting the initial polarization of the primary source to 2\%, with a polarization 
position angle parallel to the disk. The resulting intensity spectrum for both the Schwarzschild and 
Kerr cases (Fig.~\ref{Fig:Polarized_para}, top figures) is completely insensitive to the polarization 
state of the corona, such as already mentioned in the introduction. Only the polar outflows have an impact 
on the soft part of the spectra, similarly to the previous model (Fig.~\ref{Fig:Unpolarized_primary}). 
The polarized flux is, however, different since the spectra are lower in intensity than the input ones in 
the soft band. However the AGN polarized spectra become higher than the corona-disk polarized spectra at 
$E \ge$ 10~keV. Since we observe that the polarization degree is the same between the input and output 
spectra (Fig.~\ref{Fig:Polarized_para}, third panels from the top), the extra absorption brought by the 
polar outflows explains the behavior of the polarized spectra in the soft part. The initial polarization 
of the corona dominates the soft photon flux since most of the detected radiation have traveled directly 
from the source to the observer. In the case of the hard band, extra Compton scattering is responsible 
for higher final polarization degrees, 0.5 -- 1 percentage points higher at 30~keV than the input polarization, 
and thus drives the polarized fluxes to higher values. In comparison to a model with unpolarized primary photons, 
we find an overall higher polarization degree associated with the same energy-dependent variations of $\Psi$.
However, in this case, the differences between the polarization position angle of the models with and 
without parsec-scale AGN components is completely negligible.

\subsection{Type-1 spectra with a (2\% perpendicular) polarized primary source}
\label{Results:POL_PERP}

\begin{figure*}
    \includegraphics[trim = 0mm 0mm 0mm 0mm, clip, width=18cm]{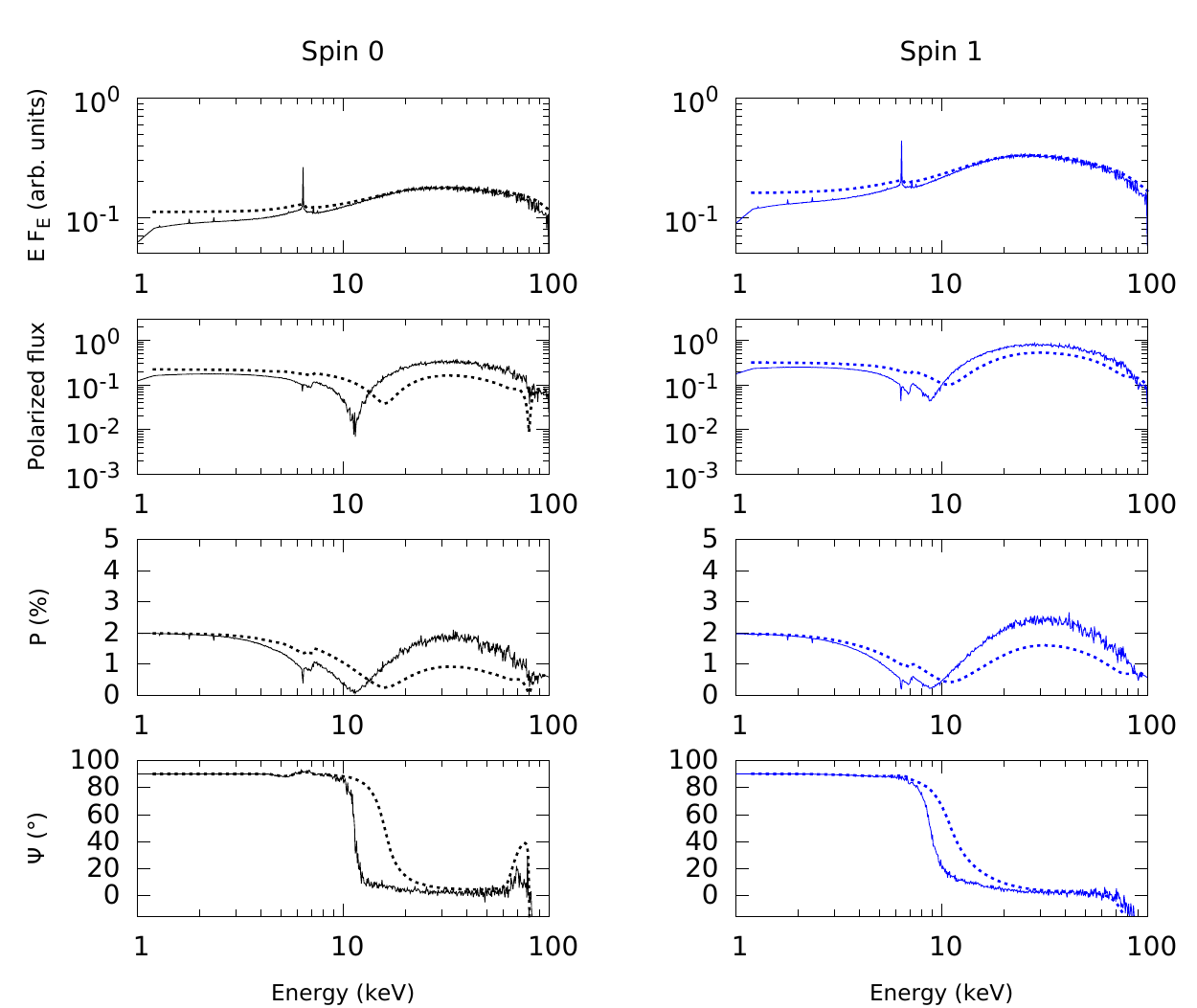}
    \caption{X-ray flux (F$_{\rm E}$ is energy flux at energy E), polarized flux, 
	    polarization degree and polarization position angle for a type-1 AGN
	    with a Compton-thick torus (n$_{\rm H_{torus}}$ = 10$^{24}$~at.cm$^{-2}$) 
	    and Compton-thin absorbing polar winds (n$_{\rm H_{wind}}$ = 10$^{21}$~at.cm$^{-2}$).
	    The input corona spectrum is 2\% perpendicularly polarized and shown in dotted line.}
    \label{Fig:Polarized_perp}
\end{figure*}

The last parametrization of the corona input polarization retains the same degree of polarization but 
changes the orientation of the polarization position angle. $\Psi$ is now perpendicular to the accretion 
disk. We observe that this different polarization setup again has no influence onto the total intensity 
spectra of both the Schwarzschild and Kerr cases (Fig.~\ref{Fig:Polarized_perp}, top figures). 
The polarized spectra (second figures from the top) are however quite different with respect to the 
previous realizations. There is a sharp dip in the polarized spectra at different energies between the 
disk-corona system (dashed line) and the full AGN model (solid line). The differences are due to the 
polarization angle of radiation. We see from the $\Psi$ energy dependence that there is a rotation of the 
polarization position angle, such as already shown in Sect.~\ref{Modeling:Polarization}. However, the 
energy at which the transition occurs is different: it happens at lower values, typically 10~keV. 
This variation is predominantly due to the extra scattering components. Since there is an additional flux 
coming from scattering onto the equatorial torus (that gives rise to parallel polarization), the amount of 
parallel polarization is larger in the case of a full AGN model than for the isolated central engine. The 
competition between parallel and perpendicular polarization is stronger and has the effect of moving the 
energy at which the transition occurs to lower values. The transition is also sharper. The dominance of 
perpendicular radiation from the primary source in the soft band explains why the final polarization
is lower than the 2\% input polarization. However, once $\Psi$ has rotated, the final degree of polarization 
increases to higher values than the initial spectra. We also find a 0.5 -- 1 percentage points increase of 
$P$ in the Compton hump, similarly to the previous cases.

%---------------------------------------------------------------------------------------------------%
\section{Discussion}
\label{Discussion}
It appears that if the type-1 intensity spectrum of AGN is affected by the presence of an equatorial torus 
and polar outflows, the former imprinting the spectrum with strong emission lines and the latter absorbing 
soft X-ray photons, it is completely insensitive to the polarization of the primary source. This conclusion 
is not new, since many authors have explored the effect of parsec-scale AGN components onto the X-ray intensity 
spectrum of AGN \citep[e.g.,][]{Ghisellini1994,Schurch2007,Schurch2009,Ricci2014}. What is innovative, is that 
the impact of those parsec-scale media have been demonstrated to alter the expected polarization signal 
from the disk-corona system even in type-1 orientations. The presence of additional scattering targets that 
are not within the influence zone of the central potential well may change the expected degree of polarization 
in different ways. If the X-ray photons produced in the hot corona are unpolarized or polarized parallelly to 
the accretion disk, we expect higher polarization degrees due to higher probabilities of scattering events. 
In particular the Compton hump shows a polarization degree on average 0.5 -- 1 percentage points higher than expected. 
The soft band ($E \le$ 10~keV) sees its polarization degree changing by a fraction of a percent, depending on the
input corona polarization degree. The polarization angle is not affected. In the case of a source of perpendicularly 
polarized photons, we find that simulations that only account for an isolated disk-corona system
usually over-estimate the true polarization degree in the soft band and under-estimate it in the Compton 
hump. Additionally, the rotation of the polarization angle between the soft and the high energy band is 
shifted towards lower energies due to the enhanced production of parallel photons by the torus.  

We have thus proven that parsec-scale AGN components are important when estimating the true X-ray polarization 
we expect from type-1 radio-quiet AGN. Varying the parametrization of the disk-corona system, such as the height 
of the corona or its initial polarization degree, will change the polarization from the central engine but not 
the impact of the distant components. First-order rescaling of the results is possible in those cases. 
On the other hand, polarization is sensitive to the inclination of the system, together with the 
geometry of the different components. It is then important to run new simulations in order to estimate 
the X-ray polarization from a complete AGN model that is viewed at a different orientation or that is 
modeled using different geometrical components. Finally, it is still unclear if varying the Compton-thickness
of the torus and winds could drastically change our results. We thus examine this question in the following subsections.

\subsection{Impact of torus thickness}
\label{Discussion:torus}

\begin{figure}
    \includegraphics[trim = 0mm 0mm 0mm 0mm, clip, width=8.5cm]{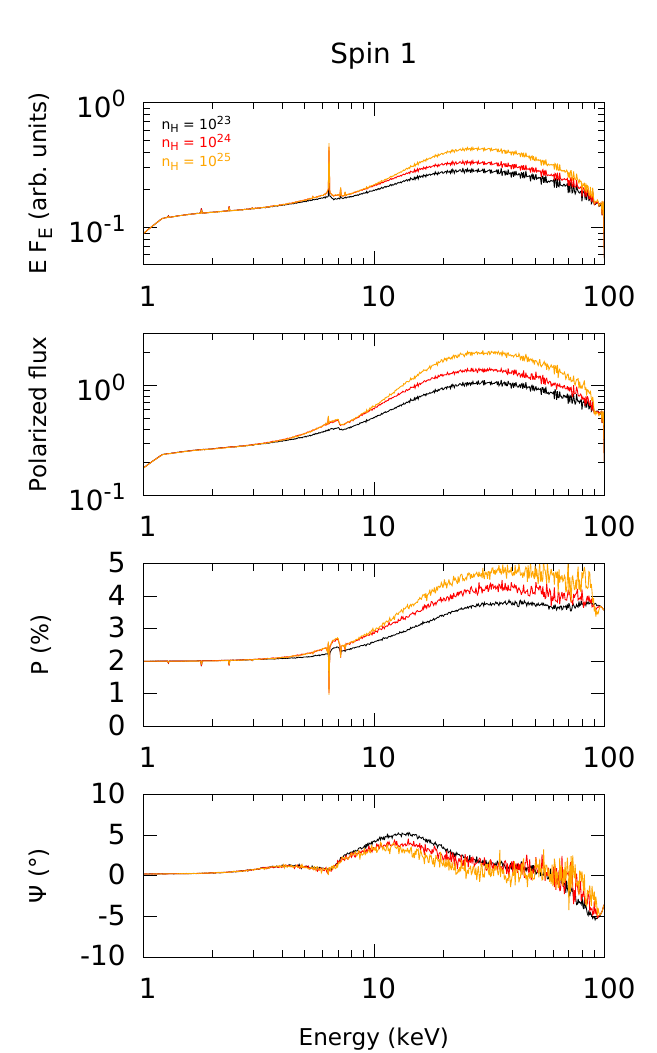}
    \caption{X-ray flux (F$_{\rm E}$ is energy flux at energy E), polarized flux, 
	    polarization degree and polarization position angle for three type-1 AGN
	    models with different torus Compton-thickness: n$_{\rm H_{torus}}$ = 10$^{23}$~at.cm$^{-2}$
	    (black), 10$^{24}$~at.cm$^{-2}$ (red), and 10$^{25}$~at.cm$^{-2}$ (orange).
	    The polar winds are Compton-thin (n$_{\rm H_{wind}}$ = 10$^{21}$~at.cm$^{-2}$) 
	    and the input corona spectrum is 2\% parallelly polarized.}
    \label{Fig:Torus}
\end{figure}

In Fig.~\ref{Fig:Torus}, we present the same AGN model as in Sect.~\ref{Fig:Polarized_para}, i.e., with a 2\%
parallel polarization primary source on top of a maximally rotating SMBH (spin 1), but we vary the hydrogen 
column density of the circumnuclear region. We examine the Compton-thin hypothesis (n$_{\rm H_{torus}}$ = 
10$^{23}$~at.cm$^{-2}$, black line), the transition value between Compton-thin and Compton-thick material 
(n$_{\rm H_{torus}}$ = 10$^{24}$~at.cm$^{-2}$, red line), and the Compton-thick case (n$_{\rm H_{torus}}$ 
= 10$^{25}$~at.cm$^{-2}$, orange line). We confirm the findings from \citet{Murphy2009,Murphy2011}: the 
reprocessed Compton hump begins to appear for column densities of 10$^{24}$~at.cm$^{-2}$ and higher. This
has an impact on the intensity and polarization spectra, in the sense that the higher densities induce a 
larger number of scattering events, hence an increase of the polarization in the hard X-ray band. However, 
the enhanced number of scattering implies a photon polarization angle that aligns more with a perfectly 
parallel orientation. This is why the energy-dependent variations of the polarization angle at high energies
tend to be smoothed out. The soft X-ray band is less affected since photo-absorption is stronger than 
scattering probabilities. 

We thus see that the hard X-ray ($E \ge$ 10~keV) polarization of type-1 Seyfert galaxies is influenced by 
parsec-scale equatorial regions. The polarization degree will be higher for higher hydrogen column
densities but the disk-corona system signatures will be ultimately washed out by Compton-thick material
along the equatorial plane. The soft X-ray band is less affected by the amount of molecular gas in the torus.
In principle it could be feasible to determine the equivalent equatorial hydrogen density by measuring the 
X-ray polarization in the soft and hard bands, given the fact that the model is not too degenerated and that
the polarimetric instrument is sensitive enough.

\subsection{Impact of wind thickness}
\label{Discussion:wind}

\begin{figure}
    \includegraphics[trim = 0mm 0mm 0mm 0mm, clip, width=8.5cm]{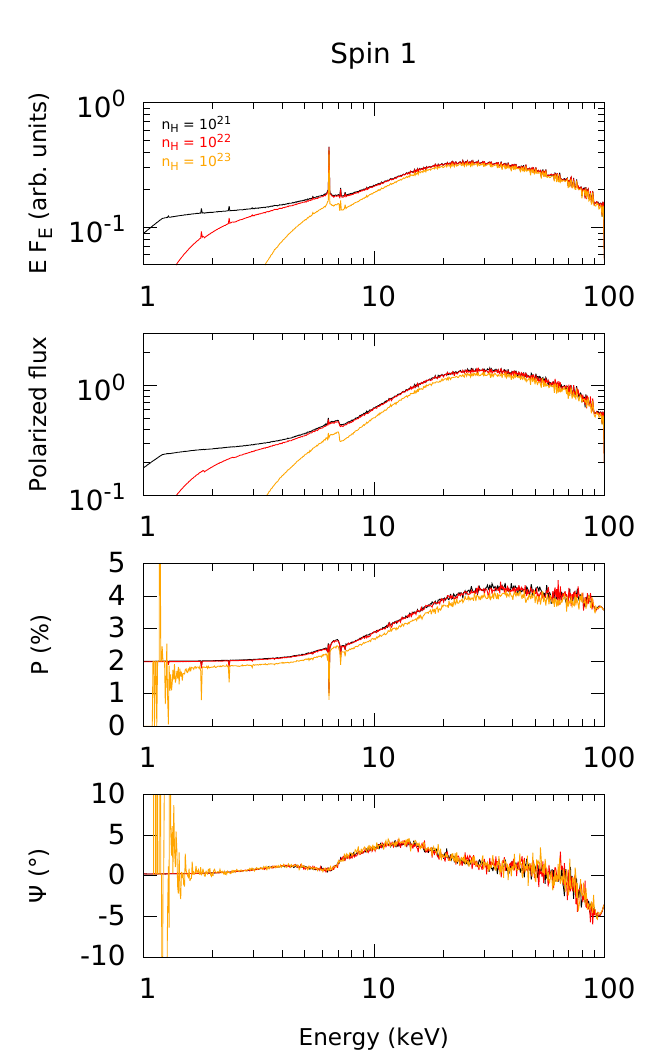}
    \caption{X-ray flux (F$_{\rm E}$ is energy flux at energy E), polarized flux, 
	    polarization degree and polarization position angle for three type-1 AGN
	    models with different wind Compton-thickness: n$_{\rm H_{wind}}$ = 10$^{21}$~at.cm$^{-2}$
	    (black), 10$^{22}$~at.cm$^{-2}$ (red), and 10$^{23}$~at.cm$^{-2}$ (orange).
	    The torus is Compton-thick (n$_{\rm H_{torus}}$ = 10$^{24}$~at.cm$^{-2}$) 
	    and the input corona spectrum is 2\% parallelly polarized.}
    \label{Fig:Wind}
\end{figure}

The second potential impacting parameter is the hydrogen column density along the observer's line-of-sight.
This is represented in the form of the polar outflow in our AGN model and we disregard additional Galactic 
absorption since it will only contribute to decrease the observed flux (forward scattering is negligible).
We thus vary the Compton-thickness of the outflows in the model presented in Sect.~\ref{Fig:Polarized_para}.
In Fig.~\ref{Fig:Wind} we test three column densities: n$_{\rm H_{wind}}$ = 10$^{21}$~at.cm$^{-2}$
(black line), 10$^{22}$~at.cm$^{-2}$ (red line), and 10$^{23}$~at.cm$^{-2}$ (orange line). This is the typical 
range of X-ray column density for Seyfert galaxies as indexed by \citet{Jimenez2008} and \citet{Fischer2014}, 
the only exceptions being the uncommon windless, bare, AGN \citep{Reeves2016,Porquet2018}. We note that, for 
the case of Ark~120, the warm absorber region is seen in emission \citep{Reeves2016}, which indicates that the 
bi-conical configuration we use is a simplification and that the geometry of the polar component might be more 
complex, including the presence of hollow winds and clumps. These different configurations have already been 
addressed in \citet{Marin2013} and \citet{Marin2015}, respectively. In the case of uniformly-filled conical winds, 
we observe that the higher the column density, the higher the absorbed flux in intensity. Additional molecular 
material leads to higher absorption probabilities \citep{Odaka2016}, but the impact on the polarization spectrum 
only appears for hydrogen column densities above 10$^{23}$~at.cm$^{-2}$. The final polarization degree 
decreases due to two cumulative effects. First, fewer photons reach the observer, being scattered away from the 
observer's line-of-sight, resulting in strong statistical noise below 2~keV. Second, a non-negligible fraction 
of photons that were out of the line-of-sight start to scatter towards the observer when the hydrogen column 
density becomes high enough to collimate radiation in a sort-of symmetric wind. The resulting photons acquire a 
random polarization, eventually contributing to the decrease of the final polarization degree we observe. 
Regarding $\Psi$, with the exception of numerical noise, the polarization position angle spectrum appears 
non-altered since transmission remains the dominant mechanism. 

We conclude that varying the amount of gas in the polar direction has a deeper impact on the total flux spectrum 
than on the polarized component of light. The higher the X-ray column density the lower the resulting polarization
degree, but this diminution is almost constant in energy and is of the order of 0.1 percentage point for the cases 
studied here. The polarization position angle is completely unaffected.

%---------------------------------------------------------------------------------------------------%
\section{Conclusions}
\label{Conclusions}

We have tested whether old simulations of type-1 AGN that only considered the disk-corona system were reliable 
for predicting the X-ray polarization of type-1 radio-quiet AGN. We found that they do give a correct estimation 
of the order of magnitude of expected polarization but the addition of parsec-scale components, such as the equatorial
circumnuclear region or polar outflows, have an impact. Depending on the initial polarization state of the 
X-ray photons produced in the corona, different outcomes may occur:

\begin{itemize}
\item If the initial photons are unpolarized or polarized parallelly to the accretion disk. In that case the polarization 
degree is almost unaffected below 3~keV, $P$ increases by 0.1 -- 0.5 percentage points between 3~keV and 10~keV, and the 
polarization degree is 0.5 -- 1 percentage points higher at higher energies. The dominant contributor to those changes 
is the parsec-scale torus that adds a parallelly polarized component to the polarization of the central engine. The final
polarization position angle is not strongly affected, but the energy-dependent variations of $\Psi$ are smoothed out
by the additional equatorial scattering contribution.\\
\item If the initial photons are polarized perpendicularly to the accretion disk. In that case the true polarization
we expect to observe in type-1 AGN is lower by 0.1 -- 0.5 percentage points between 3~keV and 10~keV with respect to what
was predicted before when the central engine was isolated. Around 10~keV the polarization position angle rotates by 90$^\circ$ 
but at a slightly lower energy than in previous estimations. Finally, at $E \ge$ 10~keV, the total polarization degree is 0.5 -- 1 
percentage points higher than predicted. All these differences are due to the torus scattered component that adds some 
polarized flux at all energies with parallel polarization. The torus polarization adds to the central engine polarization 
above 10~keV, decreases polarization below 10keV and shifts the transition of the polarization position angle to lower energies.
\end{itemize}

Overall it is possible to correct previous simulations that did not account for extra parsec-scale AGN regions
by simply increasing or decreasing the predicted degree of polarization. The polarization position angle is less affected,
except regarding the energy at which the rotation of $\Psi$ is expected if the corona produces perpendicularly polarized
photons.

Our simulations are encouraging for the future of X-ray polarimetry since they predict slightly higher polarization degree 
than expected. This means that the amount of time needed to detect the polarization threshold will be lower for several AGN. 
As an example, in the case of NGC~3783 whose X-ray flux in the 2--10~keV band varies in the range 
4--9 $\times$ 10$^{-11}$ ergs.cm$^{-2}$.s$^{-1}$ \citep{Kaspi2001}, previous simulations would have required a 1.8~Ms 
observation for a XIPE-like S-class mission \citep{Marin2012a} while, accounting for reprocessing on parsec-scale AGN 
components, the new estimation is slightly less than 1.2~Ms. We also see that detecting AGN polarization will be 
slightly easier than expected in the hard X-ray band due to enhanced Compton scattering. We thus strongly advocate 
for future X-ray polarimetric missions targeting both the soft and hard bands.

%---------------------------------------------------------------------------------------------------%
\section*{Acknowledgments}
The authors would like the anonymous referee for usefull comments and suggestions. We are also grateful to 
Delphine Porquet for her numerous comments that helped to clarify and improve the text. This research has been supported
by the Centre national d'\'etudes spatiales (CNES) thanks to the post-doctoral grant ``Probing the geometry and physics
of active galactic nuclei with ultraviolet and X-ray polarized radiative transfer''. MD thanks for the support to 
MEYS INTER-INFORM LTI17018 project.

\label{lastpage}
\clearpage

\end{document}